\documentclass[english,aps,prb, superscriptaddress,twocolumn]{revtex4}
\pdfoutput=1
\usepackage{berasans}
\usepackage[latin9]{inputenc}
\usepackage{textcomp}
\usepackage{graphicx}
\usepackage{subscript}

\makeatletter
\@ifundefined{textcolor}{}
{%
 \definecolor{BLACK}{gray}{0}
 \definecolor{WHITE}{gray}{1}
 \definecolor{RED}{rgb}{1,0,0}
 \definecolor{GREEN}{rgb}{0,1,0}
 \definecolor{BLUE}{rgb}{0,0,1}
 \definecolor{CYAN}{cmyk}{1,0,0,0}
 \definecolor{MAGENTA}{cmyk}{0,1,0,0}
 \definecolor{YELLOW}{cmyk}{0,0,1,0}
}

\usepackage{upgreek}
\usepackage{hyperref}

\makeatother

\usepackage{babel}
\begin{document}

\title{Ultraviolet and visible range plasmonics of a topological insulator}

\author{Jun-Yu Ou}

\affiliation{Optoelectronics Research Centre \& Centre for Photonic Metamaterials,
University of Southampton, UK}

\author{Jin-Kyu So}

\email{js1m10@orc.soton.ac.uk}

\selectlanguage{english}%

\affiliation{Optoelectronics Research Centre \& Centre for Photonic Metamaterials,
University of Southampton, UK}

\author{Giorgio Adamo}

\affiliation{Centre for Disruptive Photonic Technologies, Nanyang Technological
University, Singapore 637371, Singapore}

\author{Azat Sulaev}

\affiliation{School of Physical and Mathematical Sciences, Nanyang Technological
University, 637371, Singapore}

\author{Lan Wang}

\affiliation{School of Physical and Mathematical Sciences, Nanyang Technological
University, 637371, Singapore}

\author{Nikolay I. Zheludev}

\affiliation{Optoelectronics Research Centre \& Centre for Photonic Metamaterials,
University of Southampton, UK}

\affiliation{Centre for Disruptive Photonic Technologies, Nanyang Technological
University, Singapore 637371, Singapore}
\begin{abstract}
The development of metamaterials, data processing circuits and sensors
for the visible and UV parts of the spectrum is hampered by the lack
of low-loss media supporting plasmonic excitations and drives the
intense search for plasmonic materials beyond noble metals. By studying
plasmonic nanostructures fabricated on the surface of topological
insulator Bi\textsubscript{1.5}Sb\textsubscript{0.5}Te\textsubscript{1.8}Se\textsubscript{1.2}
we found that it is orders of magnitude better plasmonic material
than gold and silver in the blue-UV range. Metamaterial fabricated
from Bi\textsubscript{1.5}Sb\textsubscript{0.5}Te\textsubscript{1.8}Se\textsubscript{1.2}
show plasmonic resonances from 350 nm to 550 nm while surface gratings
exhibit cathodoluminescent peaks from 230 nm to 1050 nm. The negative
permittivity underpinning plasmonic response is attributed to the
combination of bulk interband transitions and surface contribution
of the topologically protected states. The importance of our result
is in the identification of new mechanisms of negative permittivity
in semiconductors where visible-range plasmonics can be directly integrated
with electronics.
\end{abstract}
\maketitle
\twocolumngrid

Plasmons, coupled excitations of electrons in solids and electromagnetic
fields, are responsible for brilliant colors of Roman vases and medieval
church vitrages delivered by the colloidal suspension of gold particles
in glass\cite{1freestone2007lycurgus}. They are the key to nanophotonic
applications and form responses of nanostructured metal surfaces and
artificial metamaterial photonic structures. Plasmons can localize
electromagnetic energy in the nanoscale which is crucial for the next
generation of ultra-high density optically assisted magnetic data
storage technology\cite{2kirilyuk2010ultrafast,3tsiatmas2013optical}.
Plasmons are exploited for enhancement of light-harvesting applications,
in particular photovoltaics\cite{4atwater2010plasmonics}. Plasmon
resonances are used in biological sensors such as the pregnancy test\cite{5attridge1991sensitivity}.
Plasmon-polaritons propagating on the surface are seen as a promising
information carrier for ultra-compact inter-chip interconnects applications\cite{6ozbay2006plasmonics,7gramotnev2010plasmonics,8miller2009device}
and all-optical data processing chips\cite{9almeida2004all}. However,
only a narrow class of materials can support plasmons, most notably
noble metals like silver and gold. In recent years, we saw a surge
of research aiming to identify new plasmonic media and a very substantial
progress has been possible in the search and characterization of infrared
plasmonic materials, most notably conductive oxides, nitrides and
graphene\cite{10west2010searching,11naik2013alternative,12grigorenko2012graphene}.
The UV-visible part of the spectrum remains an extremely challenging
domain for plasmonics as gold and silver have losses there, while
this spectral range remains unattainable for artificially doped semiconductors
and graphene11. If a low loss plasmonic media existed for the blue-UV
part of the spectrum, this would open a plethora of important applications
that could include a long-awaited metamaterial suitable for super-lens
with resolution breaking the diffraction limit to operate at optical
frequency\cite{13pendry2000negative,14soukoulis2011past}, a sensor
sensitive to specific near-UV resonances in proteins and DNAs, or
enhanced light concentrator for even further improvement of the density
of optical and optically-assisted data storage\cite{15wuttig2007phase},
just to mention a few. In this work, using the example of Bi\textsubscript{1.5}Sb\textsubscript{0.5}Te\textsubscript{1.8}Se\textsubscript{1.2}
(BiSbTeSe) topological insulator crystal\cite{16arakane2012tunable,17xia2013indications},
we identify a new mechanism of visible and UV plasmonic response which
is a combination of surface optical conductivity residing in a nanoscale
layer of topologically protected surface states of the crystal and
the new mechanism of bulk optical conductivity related to the dispersion
created by the interband transitions in the medium. In a series of
optical and cathode-luminescence experiments with unstructured and
nanostructured surfaces of the alloy, we show that its plasmonic properties
are very well pronounced and are superior to that of metallic gold
in the spectral range from 200 nm to $\sim500\,\mbox{nm}$ and to
that of silver from 200 nm to $\sim340\,\mbox{nm}$. In particular,
on nanostructured metamaterial surfaces of BiSbTeSe we demonstrated
profound plasmonic peaks of absorption from 350 nm to 550 nm, and
on grating arrays with periods from 150 nm to 800 nm we observed peaks
of plasmonic cathodoluminescence in the wavelength range from 230
nm to 1050 nm. 

Plasmonics is often defined as electromagnetism at the interface between
dielectrics and conductive media with negative value of the real part
of dielectric permittivity $\varepsilon$ where external field E induces
displacement field D of essentially opposite direction (D =$\varepsilon$E,
Re\{$\varepsilon$\} $<$ 0 and \textendash{}Re\{$\varepsilon$\}
$>$ Im\{$\varepsilon$\}). Under such condition the interfaces can
support highly localized oscillation and confined surface waves known
as localized plasmons and surface plasmon-polariton waves\cite{18maier2007plasmonics}.
Plasmons derive their name from plasma of free electrons in metals
or heavily doped semiconductors\cite{19kim1997structural,20hamberg1986evaporated,21hoffman2007negative}
at interfaces of which they are often observed. Despite the huge success
of modern plasmonics, the limitation of noble metals as a plasmonic
material has led to the search for alternative plasmonic materials\cite{10west2010searching,11naik2013alternative}.
In general, there are two approaches of obtaining plasmonic behavior
at the desired frequencies. One is to heavily dope semiconductors
to increase their charge carrier concentration, but this approach
has been successful only up to near-IR regime due to the difficulty
in achieving the required doping level and additional losses at such
a high doping level\cite{21hoffman2007negative,22law2012mid,23frolich2011spectroscopic,24santiago2012nanopatterning,25noginov2011transparent,26naik2012demonstration}.
The other is to mix metals with non-metals or other metals. Titanium
nitride (TiN) which has been revisited recently as a promising plasmonic
material for the visible and near-IR wavelengths falls into this category\cite{10west2010searching,27naik2012titanium,28naik2011oxides,29steinmuller1994excitation,30hibbins1998surface,31cortie2010optical}.
Apart from the bulk plasmonic materials, two-dimensional plasmonic
materials such as 2D electron gas (2DEG) systems\cite{32theis1980plasmons,33van1988quantized,34ambacher1999two}
or, recently, graphene in the mid-IR regime\cite{35chen2012optical,36yan2013damping}
have been attracting a huge interest due to their unusual properties
such as the extreme field confinement\cite{37koppens2011graphene}.
Topological insulator is a material that behaves as insulator in its
interior but whose surface contains conducting states, meaning that
electrons can only move along the surface of the material. Thus it
also falls into this category of 2D plasmonic materials\cite{38sarma2009collective}
and the plasmons on this 2DEG system has been observed at terahertz
frequencies recently\cite{39di2013observation}. 

Here, we report on the plasmonic behavior of a topological insulator,
Bi\textsubscript{1.5}Sb\textsubscript{0.5}Te\textsubscript{1.8}Se\textsubscript{1.2},
in visible and UV parts of the spectrum. In contrast to the prototypical
topological insulators such as Bi\textsubscript{2}Se\textsubscript{3}
and Bi\textsubscript{2}Te\textsubscript{3}, it has a large bulk
resistivity due to its ordered Te-Bi-Se-Bi-Te quintuple layers\cite{16arakane2012tunable}.
Our Bi\textsubscript{1.5}Sb\textsubscript{0.5}Te\textsubscript{1.8}Se\textsubscript{1.2}
single crystals were synthesized by melting high-purity (99.9999\%)
Bi, Sb, Te and Se with molar ratio 1.5:0.5:1.8:1.2 at 950\textdegree{}C
in an evacuated quartz tube. The temperature was then gradually decreased
to room temperature over a span of three weeks\cite{17xia2013indications}. The BSTS single
crystal was then cleaved along the (100) family of planes to a thickness
of  $\sim0.5\,\mbox{mm}$.

\begin{figure}
\includegraphics[width=8cm]{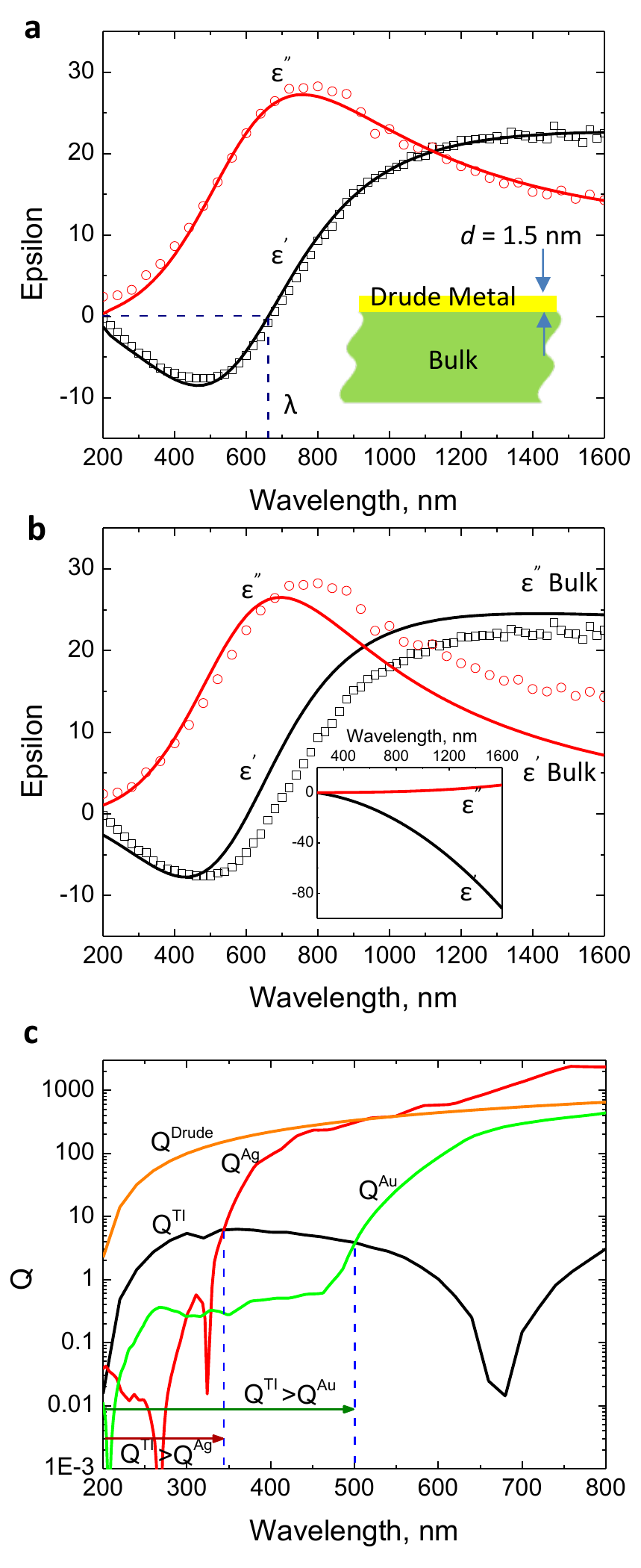}

\caption{\textbf{Plasmonic properties of the Bi\textsubscript{1.5}Sb\textsubscript{0.5}Te\textsubscript{1.8}Se\textsubscript{1.2}
topological insulator semiconductor in the visible and UV.} (a) Dielectric
function of the crystal retrieved from spectroscopic ellipsometry.
The inset shows a sketch of the layer-on-bulk model of the crystal
with a layer of topological phase of thickness d = 1.5 nm. Experimental
points are presented together with the modeling data (solid lines);
(b) An attempt to fit experimental data by the bulk contribution only
(Tauc-Lorentz dispersion formula) show the discrepancy that is attributed
to the presence of surface conductivity of topological surface state
with a Drude-like dispersion (inset). (c) Figures of merit for surface
plasmon polaritons, Q, for the Bi\textsubscript{1.5}Sb\textsubscript{0.5}Te\textsubscript{1.8}Se\textsubscript{1.2}
crystal, hypothetic Drude metal with properties of the conductive
layer on the surface of topological insulator, silver and gold. }
\label{Figure1}
\end{figure}

We first investigated the optical properties of the unstructured surface
by multiple-angle spectroscopic ellipsometry in the range from 200
nm to 1600 nm, see Fig.\ref{Figure1}. To retrieve dielectric function
from these measurements, we followed the model that was successfully
developed to represent lower frequency conductivity of this topological
insulator\cite{17xia2013indications}: we assumed a material structure
consisting of a bulk semiconductor with a thin metal film on top as
shown in the inset of Fig. \ref{Figure1}a. The semiconductor bulk
substrate supporting the thin film was modeled using the Tauc-Lorentz\cite{40jellison1996parameterization}
model that has been successfully applied to this class of materials
in the past\cite{41akrap2012optical}, while conductivity of the topological
insulator film was assumed to obey the Drude dispersion\cite{42drude1900elektronentheorie}.
From our analysis of the spectroscopic data, we found that the best-fit
parameters such as band gap of the bulk component, 0.25 eV, thickness
of the topological insulator layer, d = 1.5 nm and plasma and collision
frequencies of Drude model, 7.5 eV and 0.05 eV, respectively, are
close to previously reported values found for this material from the
independent DC conductivity measurements\cite{16arakane2012tunable,17xia2013indications}and
corroborate very well with the results of ab initio calculations of
dielectric functions of similar alloys\cite{43sharma2010first,44sharma2012electronic}.
We therefore conclude that the proposed layer-on-substrate material
structure adequately describes optical properties of Bi\textsubscript{1.5}Sb\textsubscript{0.5}Te\textsubscript{1.8}Se\textsubscript{1.2}
and use it for the interpretation of plasmonic response of the material.

Negative permittivity of Bi\textsubscript{1.5}Sb\textsubscript{0.5}Te\textsubscript{1.8}Se\textsubscript{1.2}
is clearly seen between the wavelengths, 200 \textendash{} 670 nm
(Fig. \ref{Figure1}a). It is instructive to observe that the negative
epsilon regimes in this spectral range are characteristic to both
components of the structure, the Drude layer and underlying bulk semiconductor,
as shown in Fig. \ref{Figure1}b. However the bulk contribution alone
is not sufficient to explain dielectric properties of the material
as illustrated in Fig. \ref{Figure1}b. The topologically protected
surface charge carriers are forming the response of the top metallic
layer: if such conductivity would exist in the bulk medium, plasmonic
properties of such hypothetical material would be better than that
of any known plasmonic material in this spectral range. Here we see
that even a thin layer of the topological phase has a profound impact
on plasmonic response of the structure even when placed on a lossy
substrate. Our analysis of the ellipsometry data shows that it adds
up to 8\% to the reflectivity of the nanostructured surface. 

The bulk plasmonic response here largely originates from the interband
absorption dispersion in this material, not from the bulk free carriers.
Its origin is similar to that of the negative dielectric permittivity
area at the higher frequency wing of the isolated absorption peak
that corresponds to the peak in joint density of states in case of
semiconductor. It can be understood as an inevitable consequence of
the Kramers-Kronig relations that link the real and imaginary parts
of permittivity near a strong absorption peak, as has been observed
near infrared phonon lines\cite{45spitzer1959infrared,46loh1968optical}.
Negative permittivity due to interband transitions has been theoretically
predicted for a similar material of Bi\textsubscript{2}Se\textsubscript{3}
in electronic structure calculations within the density function theory
based on full potential linearized augmented plane wave and local
orbitals\cite{43sharma2010first,44sharma2012electronic}. However,
to our knowledge Bi\textsubscript{1.5}Sb\textsubscript{0.5}Te\textsubscript{1.8}Se\textsubscript{1.2}
represents the first example of material where negative permittivity
due to interband electronic absorption is seen at the optical frequencies
experimentally. We argue that negative permittivity due to the strong
interband absorption can only be seen in semiconductors where background,
zero-frequency permittivity is small enough to be overcome by negative
resonant contribution and Bi\textsubscript{1.5}Sb\textsubscript{0.5}Te\textsubscript{1.8}Se\textsubscript{1.2}
is the first known example. 

To evaluate the competitiveness of Bi\textsubscript{1.5}Sb\textsubscript{0.5}Te\textsubscript{1.8}Se\textsubscript{1.2} as
a plasmonic material at the optical frequencies we computed the quality
factors for surface plasmon polaritons, Q\cite{10west2010searching}. Fig. \ref{Figure1}c
shows the plasmonic quality factors for the Bi\textsubscript{1.5}Sb\textsubscript{0.5}Te\textsubscript{1.8}Se\textsubscript{1.2}
crystal (Fig. \ref{Figure1}a) and widely accepted data for noble
metals\cite{47johnson1972optical,48palik1998handbook}. From there
we argue that the topological insulator is a better plasmonic material
than silver and gold in 200 \textendash{} 350 nm and 200 \textendash{}
500 nm, respectively. Good quality factors for plasmonic resonances
could be expected in nanostructures fabricated from Bi\textsubscript{1.5}Sb\textsubscript{0.5}Te\textsubscript{1.8}Se\textsubscript{1.2}
in the range from 250 to 600 nm and this is what we have illustrated
further. 

\begin{figure}
\includegraphics[width=8cm]{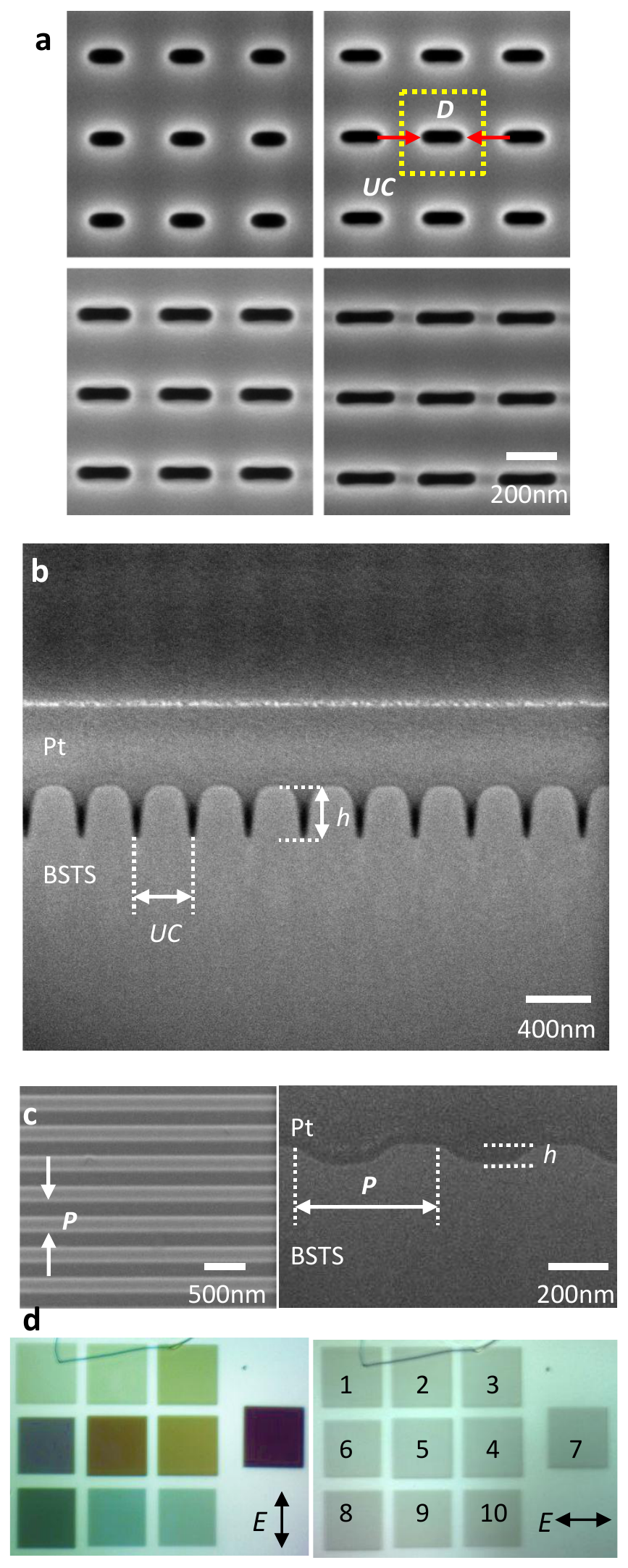}

\caption{\textbf{Nanostructured topological insulator: metamaterial and grating.}
(a) SEM images of the nano-slit array metamaterials with nominal lengths
of the slits of D = 125 nm, 150 nm, 200 nm, and 225 nm. The unit cell
(UC) size is 300 nm x 300 nm; (b) Cross-section of the array perpendicular
to the nanoslit shows the slit profile with depth of h = 300 nm. Platinum
layer was deposited for protection during sectioning; (c) SEM image
top view (left) and cross section view (right) of a grating with period
P = 500 nm and depth h = 70 nm; (d) Plasmonic colors can be seen in
optical images of nanoslits array with polarized light illumination
perpendicular to the slits and are not seen for parallel polarization.
Light\textquoteright{}s polarization is indicated by arrow. Individual
sample have sizes of 40 \textmu{}m x 40 \textmu{}m. The number 1 to
10 on right figure annotate lengths of the nanoslits in the arrays:
1 \ensuremath{\leftrightarrow} 100 nm, 2 \ensuremath{\leftrightarrow}
125 nm, 3 \ensuremath{\leftrightarrow} 150 nm, 4 \ensuremath{\leftrightarrow}
175 nm, 5 \ensuremath{\leftrightarrow} 200 nm, 6 \ensuremath{\leftrightarrow}
225 nm, 7 \ensuremath{\leftrightarrow} 250 nm, 8 \ensuremath{\leftrightarrow}
300 nm, 9 \ensuremath{\leftrightarrow} 350 nm, and 10 \ensuremath{\leftrightarrow}
400 nm, respectively. }

\label{Figure2}
\end{figure}

To verify the plasmonic behavior of Bi\textsubscript{1.5}Sb\textsubscript{0.5}Te\textsubscript{1.8}Se\textsubscript{1.2}
in nanostructures we manufactured a series metamaterials, nano-slit
antenna arrays with linear grooves cut into the surface of the crystal
and gratings on the crystal surface using focused-ion-beam milling
(Fig. \ref{Figure2}). In the nano-slit antenna array the slit length
D was varied from 100 nm to 400 nm and the period of the slit (unit
cell size, UC) was kept at 300 nm for D = 100 \textendash{} 225 nm
and UC = 1.5D for D = 250 \textendash{} 400 nm. The fabricated nano-slits\textquoteright{}
profile is close to V-shape, as shown in Fig. \ref{Figure2}b. 

\begin{figure}
\includegraphics[width=16cm]{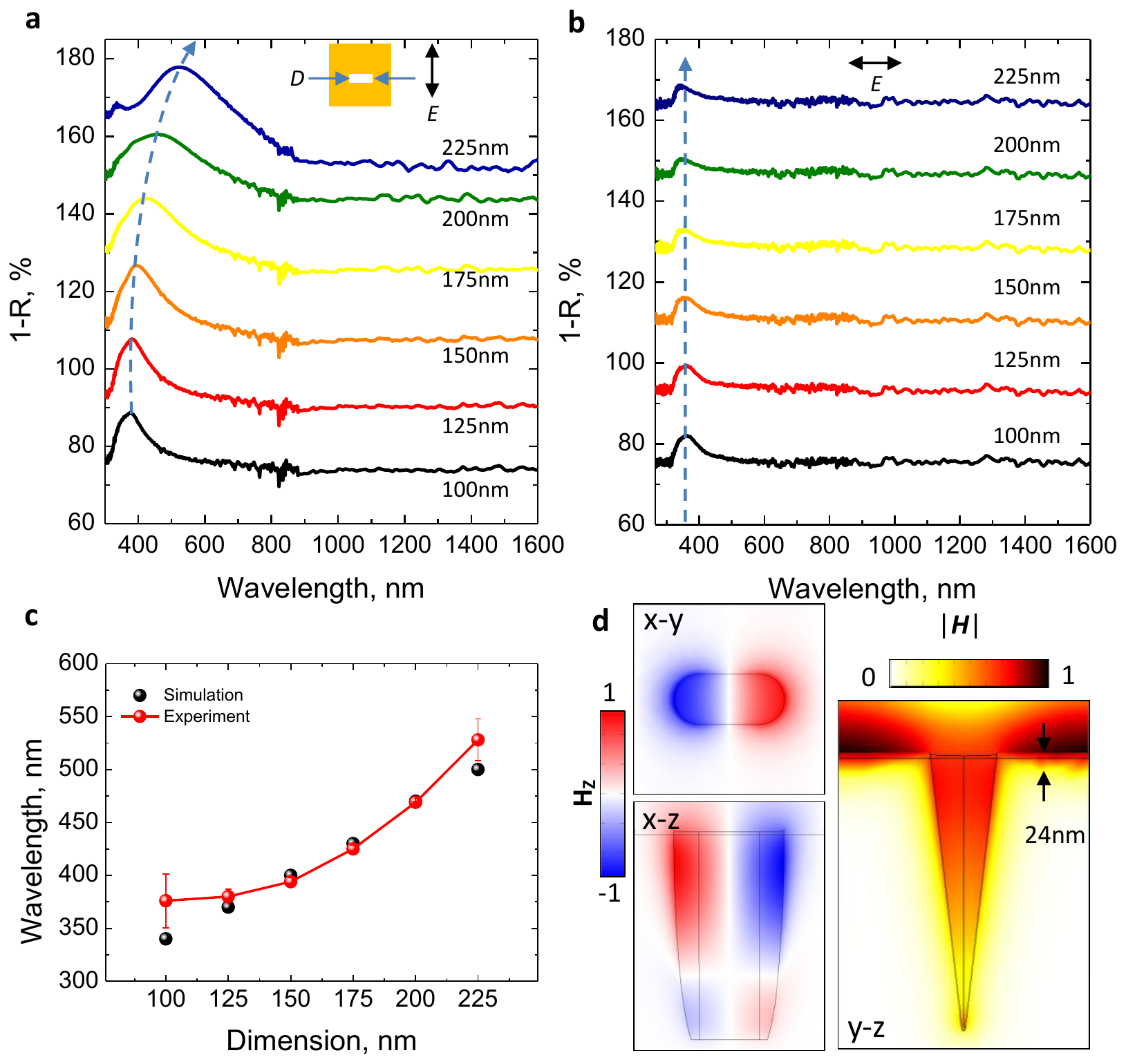}

\caption{\textbf{Plasmonic properties of nano-slit array metamaterials fabricated
on topological insulator semiconductor.} (a) Absorption spectra, 1-R,
of various nano-slit arrays with lengths D from 100 nm to 225 nm for
light polarized perpendicular to the slits nano-slits and (b) parallel
to nano-slits; (c) Peak wavelength of the absorption resonance for
various slit lengths increases with the slit length D; (d) Maxwell
modeling of the Hz- and $|$H$|$- field distribution in the slit
length of 175nm at the resonance wavelength of 430 nm indicates that
light is localized in the surface layer of thickness x $\sim24\,\mbox{nm}$ (incident light polarization is perpendicular to the nano-slit).
Different cross-sections of the slit are shown.}

\label{Figure3}
\end{figure}

The plasmonic response of the fabricated nano-slit metamaterials and
gratings were studied by measuring their reflection spectra R($\lambda$)
and their corresponding absorption spectra A($\lambda$)= 1\textendash{}R($\lambda$)
for two incident polarizations perpendicular and parallel to the nano-slits
(Figures \ref{Figure3}a and \ref{Figure3}b). The nano-slit arrays
clearly exhibited plasmonic colors for light polarized perpendicularly
to the slits, as can be seen in the optical microscope images (see
Fig. \ref{Figure2}d). A profound resonance in plasmonic absorption
can be seen for this polarization (Fig. \ref{Figure3}a). Indeed,
if a wire dipole is resonant for light polarized along the dipole,
according to the Babinet principle, slits in the conductive surface,
the \textquotedblleft{}anti-dipole\textquotedblright{}, will be resonant
for perpendicular polarization. Here, the resonant wavelength increases
monotonously with the length of the groove as can be seen in Fig.
\ref{Figure3}c. As expected, no plasmonic resonance can be found
in the polarization along the groove (see Fig. \ref{Figure3}b). On
all graphs presented on Fig. \ref{Figure3}b, an absorption peak near
400 nm is seen. It is due to a feature in the bulk interband absorption
and is also observed from unstructured surfaces. For short grooves,
the plasmonic peaks overlaps with the interband absorption feature
and became distinct for D$>$175 nm. Full 3D Maxwell calculations
of the reflectivity spectra obtained on the basis of ellipsometry
data strongly corroborate with experimental results.

\begin{figure}
\includegraphics[width=16cm]{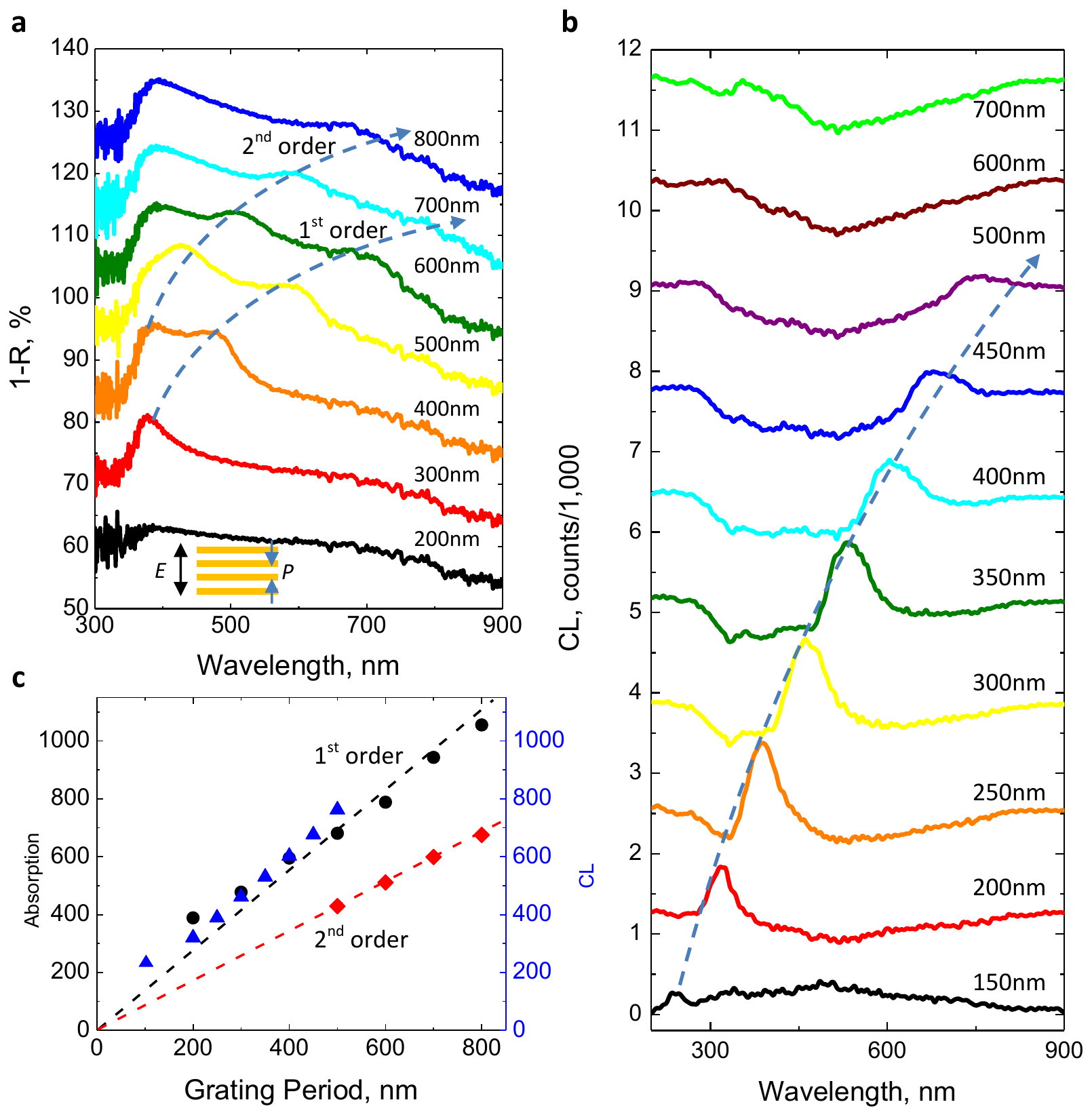}

\caption{\textbf{Plasmonic properties of gratings fabricated on topological
insulator semiconductor.} (a) Absorption spectra of one-dimensional
gratings with pitch P from 200 to 800 nm, in 100 nm steps for the
incident polarizations perpendicular to the grating; (b) Cathodoluminescence
spectra of the gratings with period from 150 nm to 700 nm; (c) Peak
wavelength of the absorption and CL resonances for various periods
of the grating (triangles- CL data; circles and rombs \textendash{}
absorption data). }

\label{Figure4}
\end{figure}

In another series of experiments, we investigated the optical response
and cathodoluminescence (CL) spectra of gratings fabricated on the
surface of topological insulator. The periods P of gratings were chosen
for their diffraction peaks to be located at different wavelengths
from the UV to visible parts of the spectrum. The width of the grating
ridge was maintained to be the half the grating period P. 70 nm deep
gratings with the periods, P from 200 nm to 1500 nm, were fabricated
on the surface of a bulk BSTS crystal as shown in Fig. \ref{Figure2}c.
The plasmonic response was found in the gratings when they were illuminated
with the polarization perpendicular to the grating rulings as shown
in Fig. \ref{Figure4}a. In contrast to the featureless absorption
spectra for the parallel polarization, the formation and evolution
of 1st and 2nd-order peaks were clearly seen for the perpendicular
polarization in the wavelengths range between 350 nm and 670 nm. Moreover,
peaks were also visible for wavelengths longer than 670 nm.

These results should be compared with our cathode-luminescence (CL)
data that have also been successfully used in the past to identify
plasmonic response\cite{49bashevoy2006generation}. The gratings were
excited with electron beam (waist diameter $\sim50\,\mbox{nm}$; electron
energy 40 keV; beam current$\sim10\,\mbox{nA}$) of a scanning electron
microscope. The electron beam was raster scanned on the area of about
10 \textmu{}m$\times$10 \textmu{}m of each grating. Figure \ref{Figure4}b
shows the normalized CL spectra from each grating where the CL from
the unstructured crystal surface was subtracted. Peaks in the range
from 230 nm to 1050 nm can be observed. The emission peaks and their
red-shift with increase of the grating period are clearly observed.
Importantly, CL peaks\textquoteright{} positions accurately match
that of the absorption peaks emphasizing their common plasmonic nature
of the responses (Fig. \ref{Figure4}c). Here, we argue that the observation
of the CL peaks for wavelengths beyond 670 nm, i.e. beyond the range
of wavelengths where dielectric permittivity of the bulk is negative,
is a clear evidence of plasmonic contribution of the surface topological
conductivity state.

In summary, we have demonstrated the plasmonic behavior of a topological
insulator, Bi\textsubscript{1.5}Sb\textsubscript{0.5}Te\textsubscript{1.8}Se\textsubscript{1.2},
at optical frequencies. It resulted from a combination of contributions
from the topologically protected surface conducting state and a strong
dispersion due to the interband transition. The optical and electron
beam excitation of the material demonstrated the existence of the
plasmonic response with quality factor of about six that is sufficient
for many sensors, light localization and metamaterials applications
and that outperforms the noble metals in the UV-blue-green parts of
the spectrum. We believe that the importance of our results is in
the identification of new class of materials with high-frequency plasmonic
response where plasmonic functionality can be directly integrated
with electronics thanks to the semiconductor nature of the material. 
\begin{acknowledgments}
The authors would like to thank Jonathan Maddock for assistance with
ellipsometry and Cesare Soci, Nikitas Papasimakis, Kevin MacDonald
and Yidong Chong for discussion. This work was supported by the Royal
Society (UK), EPSRC (UK) Programme on Nanostructured Photonic Metamaterials
and Ministry of Education Singapore, research project (Grant No. MOE2011-T3-1-005)
\end{acknowledgments}
\renewcommand{\bibnumfmt}[1]{{#1}.{\,}}
\renewcommand\refname{References}


\onecolumngrid

\end{document}